\documentstyle[aps,epsfig,floats]{revtex}

\tighten
\begin{document}

\draft

\title{
\begin{flushright}
\begin{minipage}{3 cm}
\small
hep-ph/0009343\\
VUTH 00-23
\end{minipage}
\end{flushright}
Transverse Momentum Dependence in Gluon Distribution and Fragmentation
Functions}

\author{P.J. Mulders$^{1}$ and J. Rodrigues$^{2}$}
\address{\mbox{}\\
$^1$Department of Physics and Astronomy, Free University \\
De Boelelaan 1081, NL-1081 HV Amsterdam, the Netherlands \\
\mbox{}\\
$^2$Instituto Superior T\'{e}cnico \\
Av. Rovisco Pais, 1049-001 Lisboa, Portugal\\[6mm]}

\date{\today}

\maketitle


\begin{abstract}

We investigate the twist two gluon distribution functions for spin
1/2 hadrons, emphasizing intrinsic transverse momentum of the gluons. These
functions are   relevant in leading order in the inverse hard scale in 
scattering processes such as inclusive leptoproduction or Drell-Yan scattering,
or more general in hard processes in which at least two hadrons are involved.
They show up in azimuthal asymmetries. For future estimates of such
observables, we discuss specific bounds on these functions. 

\end{abstract}





\newcommand{\be}{\begin{equation}}
\newcommand{\bea}{\begin{eqnarray}}
\newcommand{\ee}{\end{equation}}
\newcommand{\eea}{\end{eqnarray}}
\newcommand{\ba}{\begin{array}}
\newcommand{\ea}{\end{array}}
\newcommand{\bt}{\begin{tabbing}}
\newcommand{\et}{\end{tabbing}}

\newcommand{\nin}{\noindent}
\newcommand{\nn}{\nonumber}
\newcommand{\nnn}{\nonumber \\ }

\newcommand{\half}{{1 \over 2}}
\newcommand{\terco}{{1 \over 3}}
\newcommand{\quarto}{{1 \over 4}}

\newcommand{\larrow}{\leftarrow}
\newcommand{\rarrow}{\rightarrow}
\newcommand{\darrow}{\downarrow}
\newcommand{\lrarrow}{\leftrightarrow}

\newcommand{\im}{{\rm Im}}
\newcommand{\tr}{{\rm Tr}}
\newcommand{\Det}{{\rm Det}}

\newcommand{\bm}[1]{\bbox{#1}}
\newcommand{\subt}{{\scriptscriptstyle T}}
\newcommand{\subl}{{\scriptscriptstyle L}}


\newcommand{\alp}{\alpha}
\newcommand{\bet}{\beta}
\newcommand{\gamm}{\gamma}
\newcommand{\Gam}{\Gamma}
\newcommand{\Gamm}{\Gamma}
\newcommand{\del}{\delta}
\newcommand{\Del}{\Delta}
\newcommand{\eps}{\epsilon}
\newcommand{\lam}{\lambda}
\newcommand{\Lam}{\Lambda}
\newcommand{\sig}{\sigma}
\newcommand{\Sig}{\Sigma}
\newcommand{\ome}{\omega}
\newcommand{\Ome}{\Omega}

\newcommand{\mn}{\mu \nu}


\newcommand{\as}{\alpha_{\scriptscriptstyle S}}
\newcommand{\xb}{x_{\scriptscriptstyle B}}
\newcommand{\nx}{n_{\scriptscriptstyle X}}
\newcommand{\np}{n_{\scriptscriptstyle +}}
\newcommand{\nm}{n_{\scriptscriptstyle -}}


\newcommand{\bfa}{{\bf a}}
\newcommand{\bfA}{{\bf A}}
\newcommand{\bfb}{{\bf b}}
\newcommand{\bfB}{{\bf B}}
\newcommand{\bfF}{{\bf F}}
\newcommand{\bfJ}{{\bf J}}
\newcommand{\bfk}{{\bf k}}
\newcommand{\bfp}{{\bf p}}
\newcommand{\bfP}{{\bf P}}
\newcommand{\bfS}{{\bf S}}
\newcommand{\bfq}{{\bf q}}
\newcommand{\bfr}{{\bf r}}
\newcommand{\bfx}{{\bf x}}
\newcommand{\bfy}{{\bf y}}


\newcommand{\overd}{\overline{d}}
\newcommand{\overf}{\overline{f}}
\newcommand{\overF}{\overline{F}}
\newcommand{\overG}{\overline{G}}
\newcommand{\overgamm}{\overline{\gamm}}
\newcommand{\overGamm}{\overline{\Gamm}}
\newcommand{\overk}{\overline{k}}
\newcommand{\overn}{\overline{n}}
\newcommand{\overp}{\overline{p}}
\newcommand{\overP}{\overline{P}}
\newcommand{\overq}{\overline{q}}
\newcommand{\overR}{\overline{R}}
\newcommand{\overs}{\overline{s}}
\newcommand{\overS}{\overline{S}}
\newcommand{\overu}{\overline{u}}


\newcommand{\dk}{\int {d^4k \over {(2 \pi)}^4} \ }
\newcommand{\dl}{\int {d^4l \over {(2 \pi)}^4} \ }
\newcommand{\dx}{\int {d^4x \over {(2 \pi)}^4} \ }
\newcommand{\dxi}{\int {d^4 \xi \over {(2 \pi)}^4} \ }
\newcommand{\dq}{\int {d^4q \over {(2 \pi)}^4} \ }


\newcommand{\idx}{\int_0^1 dx \ }


\newcommand{\galpha}{\gamma^\alpha}
\newcommand{\gbeta}{\gamma^\beta}
\newcommand{\ggamma}{\gamma^\gamma}
\newcommand{\gmu}{\gamma^\mu}
\newcommand{\gnu}{\gamma^\nu}
\newcommand{\grho}{\gamma^\rho}
\newcommand{\gsigma}{\gamma^\sigma}

\newcommand{\gbalpha}{\gamma_\alpha}
\newcommand{\gbbeta}{\gamma_\beta}
\newcommand{\gbgamma}{\gamma_\gamma}
\newcommand{\gbmu}{\gamma_\mu}
\newcommand{\gbnu}{\gamma_\nu}
\newcommand{\gbrho}{\gamma_\rho}
\newcommand{\gbsigma}{\gamma_\sigma}

\newcommand{\ga}{\gamma_5}
\newcommand{\gz}{\gamma^0}


\newcommand{\smn}{\sig_{\mu \nu}}


\newcommand{\gab}{g^{\alpha \beta}}
\newcommand{\gag}{g^{\alpha \gamma}}
\newcommand{\gam}{g^{\alpha \mu}}
\newcommand{\gan}{g^{\alpha \nu}}
\newcommand{\gar}{g^{\alpha \rho}}
\newcommand{\gas}{g^{\alpha \sigma}}

\newcommand{\gba}{g^{\beta \alpha}}
\newcommand{\gbg}{g^{\beta \gamma}}
\newcommand{\gbm}{g^{\beta \mu}}
\newcommand{\gbn}{g^{\beta \nu}}
\newcommand{\gbr}{g^{\beta \rho}}
\newcommand{\gbs}{g^{\beta \sigma}}

\newcommand{\gga}{g^{\gamma \alpha}}
\newcommand{\ggb}{g^{\gamma \beta}}
\newcommand{\ggm}{g^{\gamma \mu}}
\newcommand{\ggn}{g^{\gamma \nu}}
\newcommand{\ggr}{g^{\gamma \rho}}
\newcommand{\ggs}{g^{\gamma \sigma}}

\newcommand{\gma}{g^{\mu \alpha}}
\newcommand{\gmb}{g^{\mu \beta}}
\newcommand{\gmg}{g^{\mu \gamma}}
\newcommand{\gmn}{g^{\mu \nu}}
\newcommand{\gmr}{g^{\mu \rho}}
\newcommand{\gms}{g^{\mu \sigma}}

\newcommand{\gna}{g^{\nu \alpha}}
\newcommand{\gnb}{g^{\nu \beta}}
\newcommand{\gng}{g^{\nu \gamma}}
\newcommand{\gnm}{g^{\nu \mu}}
\newcommand{\gnr}{g^{\nu \rho}}
\newcommand{\gns}{g^{\nu \sigma}}

\newcommand{\gra}{g^{\rho \alpha}}
\newcommand{\grb}{g^{\rho \beta}}
\newcommand{\grg}{g^{\rho \gamma}}
\newcommand{\grm}{g^{\rho \mu}}
\newcommand{\grn}{g^{\rho \nu}}
\newcommand{\grs}{g^{\rho \sigma}}

\newcommand{\gsa}{g^{\sigma \alpha}}
\newcommand{\gsb}{g^{\sigma \beta}}
\newcommand{\gsg}{g^{\sigma \gamma}}
\newcommand{\gsm}{g^{\sigma \mu}}
\newcommand{\gsn}{g^{\sigma \nu}}
\newcommand{\gsr}{g^{\sigma \rho}}


\newcommand{\hab}{g_{\alpha \beta}}
\newcommand{\hag}{g_{\alpha \gamma}}
\newcommand{\ham}{g_{\alpha \mu}}
\newcommand{\han}{g_{\alpha \nu}}
\newcommand{\har}{g_{\alpha \rho}}
\newcommand{\has}{g_{\alpha \sigma}}

\newcommand{\hba}{g_{\beta \alpha}}
\newcommand{\hbg}{g_{\beta \gamma}}
\newcommand{\hbm}{g_{\beta \mu}}
\newcommand{\hbn}{g_{\beta \nu}}
\newcommand{\hbr}{g_{\beta \rho}}
\newcommand{\hbs}{g_{\beta \sigma}}

\newcommand{\hga}{g_{\gamma \alpha}}
\newcommand{\hgb}{g_{\gamma \beta}}
\newcommand{\hgm}{g_{\gamma \mu}}
\newcommand{\hgn}{g_{\gamma \nu}}
\newcommand{\hgr}{g_{\gamma \rho}}
\newcommand{\hgs}{g_{\gamma \sigma}}

\newcommand{\hma}{g_{\mu \alpha}}
\newcommand{\hmb}{g_{\mu \beta}}
\newcommand{\hmg}{g_{\mu \gamma}}
\newcommand{\hmn}{g_{\mu \nu}}
\newcommand{\hmr}{g_{\mu \rho}}
\newcommand{\hms}{g_{\mu \sigma}}

\newcommand{\hna}{g_{\nu \alpha}}
\newcommand{\hnb}{g_{\nu \beta}}
\newcommand{\hng}{g_{\nu \gamma}}
\newcommand{\hnm}{g_{\nu \mu}}
\newcommand{\hnr}{g_{\nu \rho}}
\newcommand{\hns}{g_{\nu \sigma}}

\newcommand{\hra}{g_{\rho \alpha}}
\newcommand{\hrb}{g_{\rho \beta}}
\newcommand{\hrg}{g_{\rho \gamma}}
\newcommand{\hrm}{g_{\rho \mu}}
\newcommand{\hrn}{g_{\rho \nu}}
\newcommand{\hrs}{g_{\rho \sigma}}

\newcommand{\hsa}{g_{\sigma \alpha}}
\newcommand{\hsb}{g_{\sigma \beta}}
\newcommand{\hsg}{g_{\sigma \gamma}}
\newcommand{\hsm}{g_{\sigma \mu}}
\newcommand{\hsn}{g_{\sigma \nu}}
\newcommand{\hsr}{g_{\sigma \rho}}


\newcommand{\lc}{\epsilon^{\mu \nu \rho \sigma}}
\newcommand{\lcb}{\epsilon_{\mu \nu \rho \sigma}}


\newcommand{\km}{k_{\mu}}
\newcommand{\kn}{k_{\nu}}
\newcommand{\pmu}{p_{\mu}}
\newcommand{\pnu}{p_{\nu}}
\newcommand{\qm}{q_{\mu}}
\newcommand{\qn}{q_{\nu}}


\def\sla#1{\setbox0=\hbox{$#1$}
        \dimen0=\wd0 \setbox1=\hbox{/} \dimen1=\wd1
        \ifdim\dimen0>\dimen1 \rlap{\hbox to \dimen0{\hfil/\hfil}} #1
        \else \rlap{\hbox to \dimen1{\hfil$#1$\hfil}} / \fi}


\newcommand{\sL}{S_{\scriptscriptstyle L}}
\newcommand{\shL}{S_{h\scriptscriptstyle L}}

\newcommand{\kt}{{\bm k}_{\scriptscriptstyle T}^2}
\newcommand{\kts}{{\bm k}_{\scriptscriptstyle T}^2}
\newcommand{\ktns}{{\bm k}_{\scriptscriptstyle T}}
\newcommand{\kp}{{{\bm k}'_T}^2}
\newcommand{\kpns}{{\bm k}'_T}
\newcommand{\qt}{{\bm q}_T^2}
\newcommand{\qtns}{{\bm q}_T}
\newcommand{\pts}{{\bm p}_{\scriptscriptstyle T}^2}
\newcommand{\ptns}{{\bm p}_{\scriptscriptstyle T}}
\newcommand{\st}{{\bm S}_{\scriptscriptstyle T}^2}
\newcommand{\stns}{{\bm S}_{\scriptscriptstyle T}}
\newcommand{\stnsnb}{S_{\scriptscriptstyle T}}
\newcommand{\sht}{{\bm S}_{hT}^2}
\newcommand{\shtns}{{\bm S}_{hT}}
\newcommand{\shtnsnb}{S_{\scriptscriptstyle hT}}
\newcommand{\xpts}{(x,{\bm p}_{\scriptscriptstyle T}^2)}
\newcommand{\xkts}{(x,{\bm k}_{\scriptscriptstyle T}^2)}
\newcommand{\xktns}{(x,{\bm k}_{\scriptscriptstyle T})}
\newcommand{\xktnsps}{(x,{\bm k}_{\scriptscriptstyle T};P,S)}
\newcommand{\xktps}{(x,{\bm k}^2_{\scriptscriptstyle T};P,S)}
\newcommand{\xptns}{(x,{\bm p}_{\scriptscriptstyle T})}
\newcommand{\zkp}{(z,{\bm k}_{\scriptscriptstyle T}^{\prime 2})}
\newcommand{\zkpns}{(z,{\bm k}'_{\scriptscriptstyle T})}
\newcommand{\zkt}{(z,{\bm k}_{\scriptscriptstyle T}^2)}
\newcommand{\zktns}{(z,{\bm k}_{\scriptscriptstyle T})}

\newcommand{\ktnsnbalpb}{k_{{\scriptscriptstyle T} \alpha}}
\newcommand{\ktnsnbbetb}{k_{{\scriptscriptstyle T} \beta}}
\newcommand{\ktnsnbmu}{k_{\scriptscriptstyle T}^\mu}
\newcommand{\ktnsnbnu}{k_{\scriptscriptstyle T}^\nu}
\newcommand{\ktnsnbrho}{k_{\scriptscriptstyle T}^\rho}
\newcommand{\ktnsnbsig}{k_{\scriptscriptstyle T}^\sig}
\newcommand{\ktnsnb}{k_{\scriptscriptstyle T}}

\newcommand{\ptnsnbalp}{p_{{\scriptscriptstyle T} \alp}}
\newcommand{\ptnsnbbet}{p_{{\scriptscriptstyle T} \bet}}
\newcommand{\ptnsnbgamm}{p_{{\scriptscriptstyle T} \gamm}}
\newcommand{\stnsnbalp}{S_{{\scriptscriptstyle T} \alp}}
\newcommand{\stnsnbbet}{S_{{\scriptscriptstyle T} \bet}}
\newcommand{\stnsnbgamm}{S_{{\scriptscriptstyle T} \gamm}}

\newcommand{\ptnsnbmu}{p_{\scriptscriptstyle T}^\mu}
\newcommand{\ptnsnbnu}{p_{\scriptscriptstyle T}^\nu}
\newcommand{\ptnsnbrho}{p_{\scriptscriptstyle T}^\rho}
\newcommand{\ptnsnbsig}{p_{\scriptscriptstyle T}^\sig}
\newcommand{\ptnsnb}{p_{\scriptscriptstyle T}}

\newcommand{\stnsnbmu}{S_{\scriptscriptstyle T}^\mu}
\newcommand{\stnsnbnu}{S_{\scriptscriptstyle T}^\nu}
\newcommand{\stnsnbrho}{S_{\scriptscriptstyle T}^\rho}
\newcommand{\stnsnbsig}{S_{\scriptscriptstyle T}^\sig}


\newcommand{\gij}{g^{ij}_{\scriptscriptstyle T}}
\newcommand{\gtmn}{g^{\mu \nu}_{\scriptscriptstyle T}}
\newcommand{\gtij}{g^{ij}_{\scriptscriptstyle T}}
\newcommand{\gt}{g_{\scriptscriptstyle T}}
\newcommand{\gtbij}{g_{{\scriptscriptstyle T}ij}}
\newcommand{\epstmn}{\eps^{\mu \nu}_{\scriptscriptstyle T}}
\newcommand{\epsij}{\epsilon^{ij}_{\scriptscriptstyle T}}
\newcommand{\epstij}{\eps^{ij}_{\scriptscriptstyle T}}
\newcommand{\epstptst}{\eps^{\ptnsnb \stnsnb}_{\scriptscriptstyle T}}
\newcommand{\epstbij}{\epsilon_{{\scriptscriptstyle T}ij}}
\newcommand{\epst}{\epsilon_{\scriptscriptstyle T}}


\newcommand{\idkt}{\int d^2 \ktns}
\newcommand{\idpt}{\int d^2 \ptns}


\newcommand{\bpsi}{\overline{\psi}}
\newcommand{\dsdt}{\int [d \sigma d \tau]}
\newcommand{\dsdtau}{\int \{ d \sigma d \tau \}}
\newcommand{\langleps}{\langle P,S |}
\newcommand{\rangleps}{| P,S \rangle}
\newcommand{\moverpp}{{M \over P^+}}
\newcommand{\pdn}{P \cdot n}
\newcommand{\pp}{P^+}
\newcommand{\ppoverm}{{P^+ \over M}}
\newcommand{\ktstoverm}{{\ktns \cdot \stns \over M}}
\newcommand{\ptstoverm}{{\ptns \cdot \stns \over M}}


\section{Introduction}

Gluon distribution
and fragmentation functions are fundamental quantities in 
the study of deep inelastic scattering processes. 
In fact, together with their quark and antiquark analogues,
these process-independent quantities describe 
the soft parts of the scattering or,
in other words, the deep structure of the hadrons.
The partonic (distribution and fragmentation) functions cannot yet 
be calculated from first principles because we lack the
non-perturbative treatment of the strong interactions. 
However, valuable information on these functions can be obtained
via lattice
calculations or theoretical models.

As soon as more than one hadron is involved in a hard scattering process,
it is essential to take into account the transverse momentum  of the partons.
For instance, transverse momentum dependent quark distribution and 
fragmentation functions show up explicitly in several semi-inclusive cross
sections, in particular in azimuthal asymmetries. In the calculation of QCD
corrections for these cross sections, the inclusion of transverse momentum 
dependent gluon distributions and fragmentation functions will be necessary.
This is our motivation to study in this paper the transverse momentum
dependent gluon functions. We will follow the corresponding treatment  for
quarks developed by Mulders and Tangerman~\cite{piet95,piet96}, 
following earlier work by Ralston and Soper~\cite{ralston79}.

In a diagrammatic expansion of the hadronic tensor in powers of the
strong coupling, one finds an infinite number of gluon
correlators, which are essentially matrix elements of non-local
products of gluons fields (and sometimes quark fields) 
between hadronic states. The simplest of these
matrix elements are the ones that contain only two gluon fields.
In the $A^+ = 0$ gauge, they completely define the
twist two functions through the appropriate choice of Lorentz indices and
projections.  One has to make sure that the starting point is a gauge
invariant object,  which turns out to be a non-trivial matter. 

More complicated correlators, namely with three gluon fields, 
must also be studied. Some of these correlators will precisely provide the
link operator needed to define gauge invariant functions, and others
will reduce to gluon-gluon and gluon-quark correlators
using the QCD equations of motion. We will discard all correlators
that contribute at order $1/Q^2$ or higher in the cross sections,
$Q$ being the hard scale.

The paper is organized as follows. In the next section we start
from the gauge invariant gluon-gluon correlation function and
derive all twist two and three gluon distributions and we make an
analysis of the link  operator. We show that the link appears naturally when a
certain class of  diagrams is summed leading to a specific path. In section
\ref{twisttwo} we introduce a specific helicity basis for nucleons and 
give the spin representation of the twist two part of the gluon-gluon
correlator, which can be used to derive the natural interpretation 
of the lower twist functions as gluon densities in the framework of the 
parton model. In section \ref{frags} the
formalism is extended to the gluon  fragmentation functions, after which we
discuss some bounds. We end up with some conclusions and suggestions for
future investigations that can use the formalism developed in this paper.


\section{Gluon Correlation Functions} \label{distributions}

\begin{figure}[t]
\begin{center}
\epsfig{file=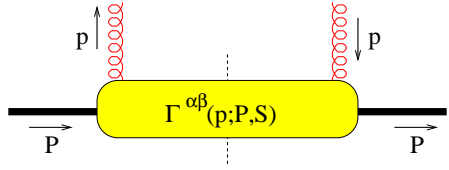,width = 6 cm}
\end{center}
\caption{\label{fig1}
The gluon correlator}
\end{figure}

In order to connect gluons in a hard scattering process to hadrons
appearing in the initial or final state, we will use (lightcone) correlation 
functions~\cite{Soper-77,Collins-Soper-82,hoodbhoy93,Bashinsky-Jaffe-98}.
Our starting point is the correlation function

\be \label{gcf1}
S^{\mu \nu}(k;P,S;n) = \dxi e^{i k \cdot \xi} 
\langle P, S | A^\mu(0) A^\nu(\xi) | P,S \rangle,
\ee

\nin 
diagrammatically represented in Fig.~\ref{fig1}.
The vectors $P$ and $S$ are respectively the momentum and the
spin of the hadron, while $k$ stands for the momentum of the gluon.
Additional dependence can come for instance from fixing the gauge using a
vector $n$. A summation over color indices is understood. If, as done
in this paper we use the
notation $A_\mu (\xi) \equiv A^a_\mu (\xi) T^a$, where the $T^a$ are the
generators of the $SU(3)$ color group, this summation is actually an
appropriate tracing.

A corresponding gauge invariant object is the quantity

\be \label{gigcf1}
\Gamm^{\mu \nu ; \rho \sig} (k;P,S) = 
\int {d^4 \xi \over (2 \pi)^4} \ e^{i k \cdot \xi} \
\langle P, S | F^{\mu \nu} (0) \ {\cal U} (0,\xi) \ F^{\rho \sig}(\xi) 
| P, S \rangle,
\ee

\nin where $F_{\mu \nu} (\xi) \equiv F_{\mu \nu}^a (\xi) T^a$ 
is the field tensor, related to the potential by 
$F_{\mu \nu} = \partial_\mu A_\nu - \partial_\nu A_\mu 
-ig\left[A_\mu,A_\nu \right]$.
The link operator ${\cal U}$ 
will be studied in detail in one of the next sections.

\subsection{The Lorentz structure of the gluon correlator}

The Lorentz structure of the gluon correlator is limited by
constraints following from hermiticity and parity conservation. These
are
\bea
\Gamm^{\rho \sig ; \mu \nu *} (k;P,S) & = & 
\Gamm^{\mu \nu ; \rho \sig} (k;P,S), \label{c1} \\
\Gamm^{\mu \nu ; \rho \sig} (k;P,S) & = & 
\Gamm_{\mu \nu ; \rho \sig} (\overk; \overP, - \overS) \label{c2},
\eea

\nin where $\overk$ = $(k^0, -k^i)$. A possible parameterization of
$\Gamm^{\mu \nu ; \rho \sig}$  compatible with these constraints is
\bea
\label{amplitudes}
\Gamm^{\mu \nu ; \rho \sig} (k;P,S) & = & 
X_1 \eps^{\mu \nu \alp \bet} \left. \eps^{\rho \sig} \right._{\alp \bet}
+ {X_2 \over M^2} 
P^{\left[ \mu \right. } g^{ \left. \nu \right] \left[ \rho \right. }
P^{ \left. \sig \right]} 
+ {X_3 \over M^2} \ k^{\left[ \mu \right. } 
g^{ \left. \nu \right] \left[ \rho \right. }
k^{ \left. \sig \right]} \nn \\
& & + {X_4 + i X_5 \over M^2} \ P^{\left[ \mu \right. } 
g^{ \left. \nu \right] \left[ \rho \right. }
k^{ \left. \sig \right]} 
+ {X_4 - i X_5 \over M^2} \  k^{\left[ \mu \right. } 
g^{ \left. \nu \right] \left[ \rho \right. }
P^{ \left. \sig \right]} \nn \\
& & + {X_6 \over M^4} \ P^{\left[ \mu \right. } k^{ \left. \nu \right]} 
P^{ \left[ \rho \right. } k^{ \left. \sig \right]} 
- {2 X_7 \over M} \ \eps^{\mu \nu \rho \sig} ( k \cdot S) \nn \\
& & + i {X_8 \over M} 
\ \eps^{\mu \nu P \left[ \sig \right. } S^{\left. \rho \right] }
+ i {X_9 \over M} \ 
\eps^{\mu \nu S \left[ \sig \right. } P^{\left. \rho \right] } \nn \\
& & + i {X_{10} \over M} 
\ \eps^{\mu \nu k \left[ \sig \right. } S^{\left. \rho \right] }
+ i {X_{11} \over M} \ 
\eps^{\mu \nu S \left[ \sig \right. } k^{\left. \rho \right] } \nn \\
& & + i {X_{12} \over M^3} \ 
\eps^{\mu \nu P \left[ \sig \right. } P^{\left. \rho \right] } (k \cdot S) 
+ i {X_{13} \over M^3} \ 
\eps^{\mu \nu k \left[ \sig \right. } k^{\left. \rho \right] } (k \cdot S) 
\nn \\
& & + i {X_{14} \over M^3} \ 
\eps^{\mu \nu P \left[ \sig \right. } k^{\left. \rho \right] } (k \cdot S) 
+ i {X_{15} \over M^3} \ 
\eps^{\mu \nu k \left[ \sig \right. } P^{\left. \rho \right] } (k \cdot S) 
\nn \\
& & + {X_{16} + i X_{17} \over M^3} \ \eps^{\mu \nu P S} 
\ k^{\left[ \rho \right.} P^{\left. \sig \right]}
+ {X_{16} - i X_{17} \over M^3} \ \eps^{\rho \sig P S} 
\ k^{\left[ \mu \right.} P^{\left. \nu \right]} \nn \\
& & + {X_{18} + i X_{19} \over M^3} \ \eps^{\mu \nu k S} 
\ k^{\left[ \rho \right.} P^{\left. \sig \right]}
+ {X_{18} - i X_{19} \over M^3} \ \eps^{\rho \sig k S} 
\ k^{\left[ \mu \right.} P^{\left. \nu \right]} \nn \\
& & + {X_{20} + i X_{21} \over M^3} \ \eps^{\mu \nu k P} 
\ P^{\left[ \rho \right.} S^{\left. \sig \right]}
+ {X_{20} - i X_{21} \over M^3} \ \eps^{\rho \sig k P} 
\ P^{\left[ \mu \right.} S^{\left. \nu \right]} \nn \\
& & + {X_{22} + i X_{23} \over M^3} \ \eps^{\mu \nu k P} 
\ k^{\left[ \rho \right.} S^{\left. \sig \right]}
+ {X_{22} - i X_{23} \over M^3} \ \eps^{\rho \sig k P} 
\ k^{\left[ \mu \right.} S^{\left. \nu \right]} \nn \\
& & + {X_{24} + i X_{25} \over M^5} \ \eps^{\mu \nu k P} 
\ k^{\left[ \rho \right.} P^{\left. \sig \right]} (k \cdot S)
+ {X_{24} - i X_{25} \over M^5} \ \eps^{\rho \sig k P} 
\ k^{\left[ \mu \right.} P^{\left. \nu \right]} (k \cdot S).
\eea

\nin The amplitudes $X_i$ as well as the original correlator
$\Gamm^{\mu \nu ; \rho \sig} (k;P,S)$ have dimensions $M^{-2}$.
With the chosen parametrization, i.e. the appropriate introduction
of factors 1 or $i$ for symmetric and antisymmetric tensors,
hermiticity in Eq. (\ref{c1}) implies that one finds real amplitudes $X_i$.
The parity constraint in Eq.~(\ref{c2})  requires that even numbers of 
$\eps$-tensors are combined only with vectors $k$ and $P$,
while odd numbers of $\eps$-tensors are combined with the
axial vector $S$ besides vectors $k,P$. Since $S$ parametrizes the nucleon
density matrix, it can only appear linearly.

Time reversal invariance, when applicable, imposes a third condition,

\be
\Gamm_{\mu \nu;\rho \sig} (\overk; \overP, \overS) 
= {\Gamm^{\mu \nu;\rho \sig}}^* (k;P,S),
\ee

\nin which implies $X_i^\ast = - X_i$ for the amplitudes
$X_5$, $X_7$,  $X_{16}$, $X_{18}$, $X_{20}$, $X_{22}$ and $X_{24}$. For this
reason, such amplitudes are called T-odd. They thus vanish when
time-reversal can be used as a constraint.

\subsection{Twist expansion}

Leading and non-leading contributions to the hadronic tensor
are easier to identify if one uses a suitable parameterization 
of the hadron momentum and spin vectors in terms of two light-like directions,
$\np$ and $\nm$ (such that $\np \cdot \nm = 1$), and two transverse 
directions. They are chosen such that the hadrons
have no transverse momentum, what means that $P$ can be written in terms of 
the light-like vectors. The momentum of the gluon and the spin vector of
the hadron must include a transverse component:
\bea
P & = & P^+ \np + {M^2 \over 2 P^+} \ \nm, \label{v1} \\
k & = & xP^+ \np + {k^2 + \kts \over 2xP^+} \ \nm + \ktnsnb, \label{v2} \\
S & = & \sL {\pp \over M} \ \np - \sL {M \over 2 \pp} \ \nm + \stnsnb 
\label{v3}.
\eea

\nin
The quantity $x$ represents the fraction of the light-cone momentum in
the $+$ direction carried by the parton. The parameter $\sL$ and the two
component  vector $\stns$ are such that $\sL^2 + \st = 1$. The quantity
$\sL$ is referred to as the helicity.
Having defined the $n_\pm$ vectors one has transverse tensors
$\gtmn$ and $\epstmn$ defined as
\bea
\gtmn & \equiv & \gmn - \np^\mu \nm^\nu - \np^\nu \nm^\mu, \\
\epstmn & \equiv & \eps^{\np \nm \mu \nu} = \eps^{-+\mu\nu}.
\eea
The expansion in lightlike vectors shows its usefulness only when the
correlation functions are used in a calculation of a hard scattering process
in which a hard timelike or spacelike vector appears setting the hard scale.
An example is the momentum transfer squared, $q^2 = - Q^2$ in
deep inelastic leptoproduction. For the soft part in the process,
described with the correlation functions it implies that after all
calculations are finished, the $\pp$ is of the same order of magnitude as 
the hard scattering scale $Q$.  
Simple power counting tells us that the most important contributions
from $\Gamm^{\mu \nu ; \rho \sig}$ 
are the ones with the largest possible number of $+$ indices.
Due to the antisymmetric character of the field tensor, these are
obviously $\Gamm^{+ i; + j}$ (referred to as twist two), 
followed by $\Gamm^{+i;+-}$ and $\Gamm^{ij ; l +}$ (referred to as
twist three),
where $i,j,l,\ldots$ indicate transverse indices.

For deep inelastic scattering processes we always need the soft parts 
integrated over the momentum component $k^-$. 
Starting with the twist two part,
we define

\be \label{l22}
M \Gamm^{ij}(x, \ktns) \equiv \int dk^- \ 
\Gamm^{+ i; + j} (k;P,S) 
= \left. \int \frac{d\xi^-d^2\xi_{\scriptscriptstyle T}}{(2\pi)^3}
\ e^{ik\cdot \xi} \,\langle P,S\vert F^{+i}(0)\,F^{+j}(\xi)\vert P,S\rangle
\right|_{\xi^+ = 0} .
\ee

\nin
This quantity depends on the momentum fraction $x$ and the transverse
momentum $\ktns$ besides the (suppressed) dependence on the target momentum 
and spin, and is conveniently expressed in terms of transverse
tensors and vectors. Concerning the dependence on the hadron spin, we
furthermore separate the unpolarized (O),
longitudinally polarized (L) and transverse polarized (T) situations.
This leads to
\bea
\label{par1}
\Gamm_O^{ij} \xktns & = & {x \over 2} \ {\pp \over M} 
\left[ - \gtij G \xkts 
+ \left( {\ktnsnb^i \ktnsnb^j \over M^2} + \gtij {\kts \over 2M^2} \right)
H^\perp \xkts \right], \\
\label{par2}
\Gamm_L^{ij} \xktns & = & {x \over 2} \ {\pp \over M} 
\left[ -i \epstij \sL \ \Del G_L \xkts
+ {\epst^{\ktnsnb \left\{ i \right. } \ktnsnb^{\left. j \right\}} \over 2M^2}
\ \sL \ \Del H_L^\perp \xkts \right], \\
\label{par3}
\Gamm_T^{ij} \xktns & = & {x \over 2} \ {\pp \over M} 
\Biggl[ - \gtij \ {\epst^{\ktnsnb \stnsnb} \over M} \ G_T \xkts
- i \epstij \ {\ktns \cdot \stns \over M} \ \Del G_T \xkts \nn \\
& & \qquad\quad 
+ {\epst^{\ktnsnb \{ i } \ktnsnb^{j \}} \over 2M^2}
\ktstoverm \ \Del H_T^\perp \xkts 
\nn \\ & & \qquad\quad 
+ {\epst^{\ktnsnb \left\{ i \right. } \stnsnb^{\left. j \right\}} +
\epst^{\stnsnb \left\{ i \right. } \ktnsnb^{\left. j \right\}} \over 4M}
\,\left[ \Del H_T \xkts - \frac{\kts}{2M^2}\,\Del H_T^\perp \xkts
\right] \Biggr],
\eea
\nin where the expressions of the functions in terms of the amplitudes $X_i$
can be found in the appendix. 
The factors in this parametrization are
chosen in order that $G$ can be interpreted as the gluon momentum
density~\cite{Collins-Soper-82}, which will become clear when we discuss sum
rules at the end of this section and in the next section.
Actually also the use of the combination
$\Delta H_T^\prime = \Delta H_T - \Delta H_T^{\perp (1)}$, where
$\Delta H_T^{\perp (n)} \equiv (\kts/2M^2)^n \Del H_T^\perp$
(with similar definitions for other functions), is done
because it is nicer for interpreting the functions.

When we use the soft parts in calculations up to ${\cal O}(1/Q)$ we need
$\Gamm^{+i;+-}$ and $\Gamm^{ij;l+}$, again 
integrated over $dk^-$, which we refer to as twist three contributions,

\bea 
M \Gamm^{i-}(x, \ktns) & \equiv & \int dk^- \ 
\Gamm^{+ i; + -} (k;P,S) ,
\\
M \Gamm^{ij,l} (x,\ktns) &\equiv & \int dk^- \ \Gamm^{ij;l+}(k;P,S) .
\eea
They are again parametrized in terms of a number of functions.
We obtain for the various hadron polarizations,
\bea
\Gamm_O^{i-} \xktns & = & {x \over 2}
{ \ktnsnb^i \over M} \ G_3^\perp \xkts , \\
\Gamm_L^{i-} \xktns & = & {x \over 2}
i \sL \ {\epst^{\ktnsnb i} \over M} \Del G_{3L}^\perp \xkts, \\
\Gamm_T^{i-} \xktns & = &
{x \over 2}  \left[
i \epst^{\stnsnb i} \ \Del G_{3T}^\prime \xkts
+ i {\epst^{\ktnsnb i} \over M} \ \ktstoverm
\ \Del G_{3T}^\perp \xkts \right]
\nn \\ & = &
{x \over 2}  \left[
i \epst^{\stnsnb i} \ \Del G_{3T} \xkts
+ i \epst^{i  \alp } \stnsnb^{\bet}
\left( {\ktnsnbalpb \ktnsnbbetb \over M^2} + g_{T \alp \bet} {\kts \over 2M^2}
\right) \Del G_{3T}^\perp \xkts \right],
\eea
and
\bea
\Gamm^{ij;l}_O (x, \ktns) & = & \ {x \over 2} \
{- \gt^{l \left[ i \right. } \ktnsnb^{\left. j \right] } \over M}
\ H_3^\perp \xkts ,
\\
\Gamm^{ij;l}_L (x, \ktns) & = &
{x \over 2} \ i \sL \epstij
\ {\ktnsnb^l \over M} \ \Del H_{3L}^\perp \xkts ,
\\
\Gamm^{ij;l}_T (x, \ktns) & = &
{x \over 2} \left[
i \epstij \stnsnb^l \ \Del H_{3T}^\prime \xkts  +
i \epstij \ {\ktnsnb^l \over M} \ \ktstoverm
\ \Del H_{3T}^\perp \xkts \right]
\nn \\ & = & 
{x \over 2} \left[
i \epstij \stnsnb^l \ \Del H_{3T} \xkts
-i \epstij \ S_{T\alp}\left({\ktnsnb^\alp \ktnsnb^l \over M^2}
+\gt^{\alp l}{\kts\over 2M^2}\right)
\ \Del H_{3T}^\perp \xkts \right]
\eea
Again the functions expressed in terms of the amplitudes are given in
the appendix. While the functions for twist two are real
functions, those for twist three are arranged in terms of complex functions in
such a way that the T-even functions correspond to the real parts and the
T-odd functions correspond to the imaginary parts. 
For twist two, the functions $G_T$, $\Delta H_L^\perp$, $\Delta H_T^\perp$, 
and $\Delta H_T$ are T-odd.

Let us make a short remark on the names of the functions, which
follow for the indices in part the notations of quark distributions as
introduced in \cite{jaffe92}  (and extended in \cite{piet95,piet96}). 
The distributions are represented by $G$, $H$, $\Delta G$ or $\Delta H$. 
The names $G$ and $\Delta G$ are reserved for functions
that do not involve uncontracted momentum indices. These do not flip the
gluon helicity, and represent unpolarized ($G$) and polarized ($\Delta G$)
gluons, respectively. The functions $H$ and $\Delta H$ flip gluon helicity
in unpolarized or polarized targets respectively, as we will discuss
in detail below. We distinguish between the longitudinally polarized spin 1/2
target being multiplied by $\sL$, which acquire a subscript $L$ and those
appearing in a transversely polarized spin 1/2 target being multiplied by the
transverse spin of the hadron, which acquire a subscript $T$. If there is an
uncontracted component of the transverse momentum of the gluon multiplying the
function or if needed to avoid double names, we
add a superscript $\perp$. Finally, twist three functions are given an
additional subscript `3'. 

In the next step we perform the integration over the transverse
momentum of the gluon to arrive at the distribution functions which 
depend only on $x$ and are important in deep inelastic inclusive measurements.
We define

\be
\Gamm^{ij} (x) \equiv  \int d^2 \ktns \ \Gamm^{ij} \xktns
= \left. \int \frac{d\xi^-}{2\pi}
\ e^{ik\cdot \xi} \,\langle P,S\vert F^{+i}(0)\,F^{+j}(\xi)\vert P,S\rangle
\right|_{\xi^+ = \xi_{\scriptscriptstyle T} = 0} ,
\label{xcor}
\ee

\nin
and similarly for $\Gamm^{i-} (x)$ and  $\Gamm^{ij;l} (x)$.

We find (combining the polarizations)
\bea \label{paula1}
\Gamm^{ij} (x) & = & {x \over 2} \ {\pp \over M} \left[ -\gtij G(x)
- \sL i \epstij \ \Del G (x) \right], \\
\Gamm^{i-} (x) & = & {x \over 2} \ i \epst^{\stnsnb i}
\ \Del G_{3T}(x), \\
\Gamm^{ij,l} (x) & = & 
{x \over 2} \ i \epst^{ij} \stnsnb^l \ \Del H_{3T}(x) .
\eea
where $G(x) = \int d^2 \ktns \ G \xkts$ 
and similarly for $\Delta G_{3T}$ and $\Delta H_{3T}$, 
while $\Delta G(x) = \int  d^2 \ktns \Delta G_L\xkts$.
The functions $\Delta G_{3T}$ and $\Delta H_{3T}$ are in essence the
functions $H_1$ and $H_2$ of Ref.~\cite{hoodbhoy93}.


\subsection{Sum rules}

Local hadronic matrix elements are obtained from the gluon 
correlation functions after integration over $dk^+$, e.g. 

\be
M^2 \Gamm^{i j}  \equiv M \int dk^+ \ \Gamm^{i j}(x)
= \langle P,S | F^{+ \mu}(0) F^{+ \nu}(0) | P, S \rangle.
\ee

The trace of this quantity  is precisely the gluon part of the
energy momentum tensor. Using the parametrization of $\Gamma^{ij}(x)$
in terms of the gluon distribution $G(x)$ one finds

\be
M^2 \Gamm^{ii} 
= 2 (\pp)^2 \int_0^1 dx \ x G(x)
= \langle P,S | \theta^{++} | P,S \rangle .
\ee

\nin To derive this last relation we used the fact that the integral over
$dx$ has a support between $-1$ and $+1$ 
and the symmetry relation $G(-x) = - G(x)$, 
which follows from the commutation relations for gluonic fields. The number

\be \label{epsg}
\eps_g = \int_0^1 dx \ x G(x)
\ee

\nin thus is identified with the fraction of light-cone 
momentum carried by the gluons, $0 \le \eps_g \le 1$.


\subsection{Gauge Invariance} \label{linkoperator}

\begin{figure}[t]
\begin{center}
\begin{minipage}{7cm}
\begin{center}
\epsfig{file=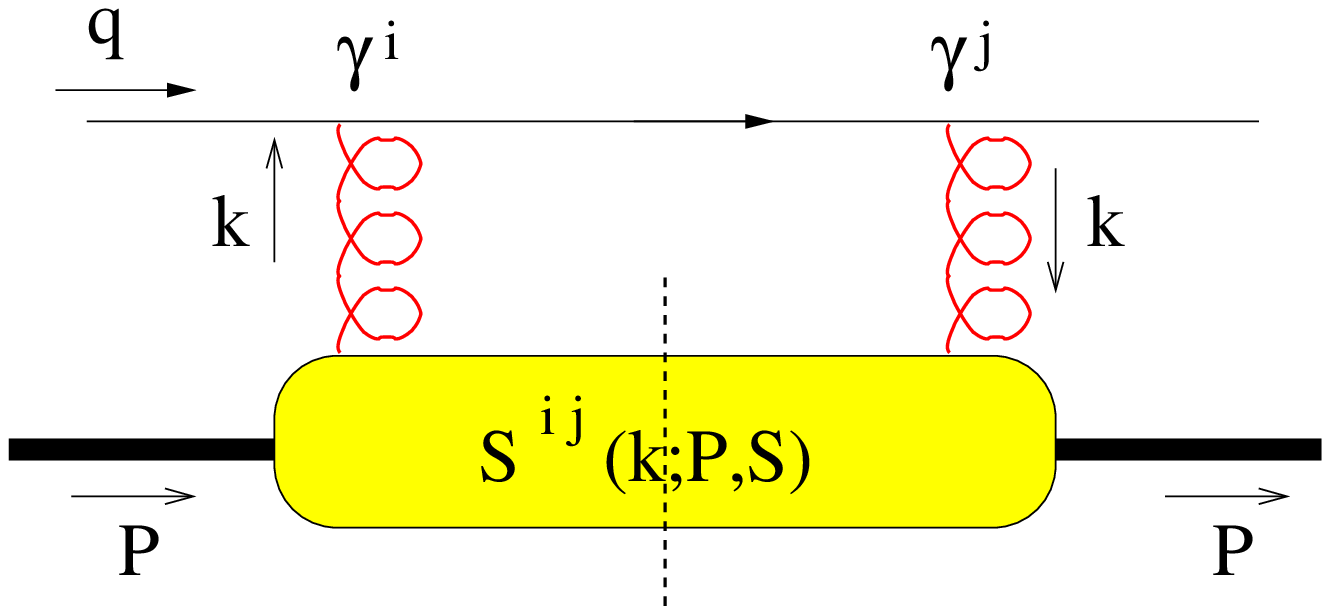,width = 6 cm}
\\
(a)
\end{center}
\end{minipage}
\hspace{2cm}
\begin{minipage}{7cm}
\begin{center}
\epsfig{file=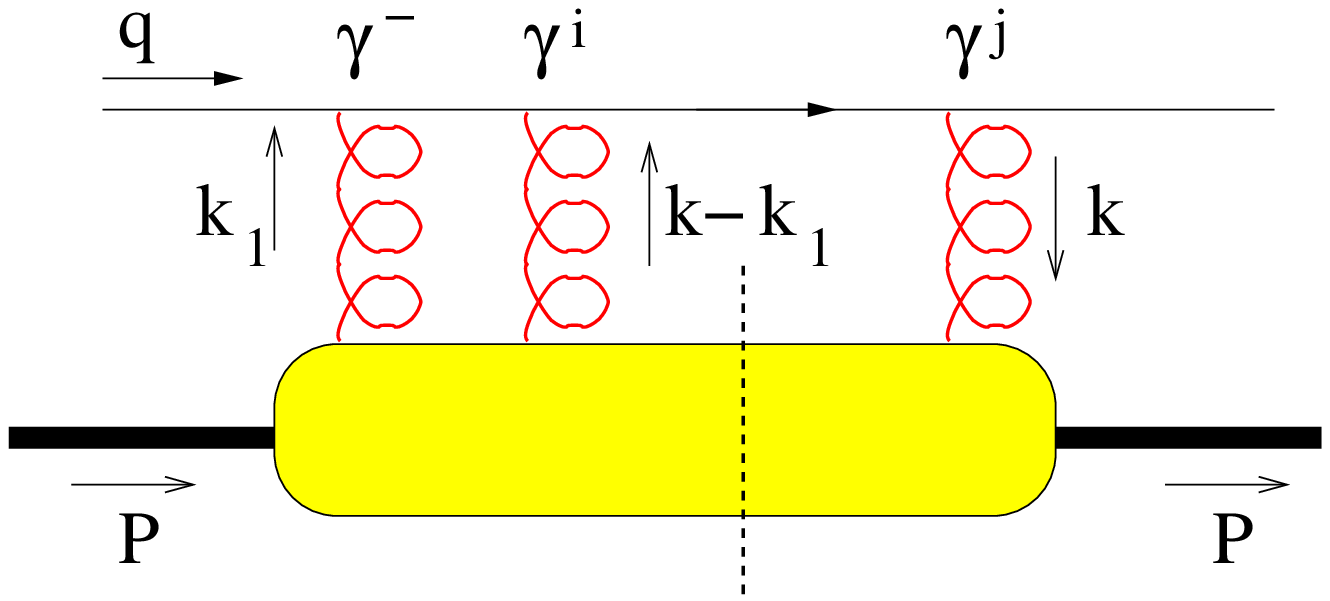,width = 6 cm}
\\
(b)
\end{center}
\end{minipage}
\end{center}
\caption{\label{fig2}
Some contributions to the fictitious highly virtual fermion-nucleon
cross section}
\end{figure}

Of course the object appearing in the diagrammatic expansion,
$S^{ij}(k;P,S)$, is not gauge invariant. We solved half of the problem
by starting with $\Gamm^{\mu\nu;\rho\sigma}$.
In particular, in the $A^+ = 0$ gauge  one has $F^{+i} = \partial^+A_T^i$
and thus

\be \label{linda}
\Gamm^{+ i ; + j} (k;P,S) = -(k^+)^2 \ S^{ij}(k;P,S).
\ee

\nin
In a general gauge,
however, one also needs to consider matrix elements of the form
$\langle A^i A^+ A^j \rangle$, $\langle A^i A^+ A^+ A^j\rangle$, etc.
Two simple leading contributions are shown in a fictitious
highly virtual fermion-nucleon scattering process in Fig.~\ref{fig2}.
These will contribute at the same order in an expansion in the inverse
hard scale. They will assure that in a general gauge one also finds
other terms such as $[A^+,A^i]$ terms in $F^{+i}$, and more importantly 
a gauge link
operator. To be precise in a fictitious calculation as in Fig.~\ref{fig2}
one finds that the field $A_T^i(\xi)$ appearing in the correlator
in Eq.~\ref{gcf1} is to be replaced by

\be 
A_T^i(\xi) = \int_\infty^{\xi^-} d\eta^-\ {\cal U}(\infty, \eta)
\,F^{+i}(\eta),
\ee
where $\eta^+ = \xi^+ = 0$, and $\eta_{\scriptscriptstyle T}$ =
$\xi_{\scriptscriptstyle T}$ and
\be
{\cal U}(\infty,\xi) = {\cal P} 
\ \exp \left( -i g \int_\infty^\xi ds^\mu A_\mu(s) \right) ,
\ee

\nin
where the path runs along the minus direction from the point
$\xi^- = \infty$ to $\xi^-$ with $\xi^+$ = 0 and $\xi_{\scriptscriptstyle T}$ 
fixed, analogous to the path for quark correlators~\cite{bm00}.
In the $k_{\scriptscriptstyle T}$ integrated correlators both links run 
along $\xi_{\scriptscriptstyle T}$ = 0
and a straight link between the lightlike separated points in the 
$\langle F^{+i}(0)\,F^{+j}(\xi)\rangle$ correlator in Eq.~\ref{xcor}
remains. For the non-integrated correlator in Eq.~\ref{l22} the
links do not close, but with the physical assumption that hadronic
matrix elements of the type $\langle A_T^i(0)A_T^j(\eta^- = \infty)
A_T^j(\xi^-)\rangle$ vanish this does not pose a problem. Furthermore
when considering weighted $k_{\scriptscriptstyle T}$-integrated cross sections
as was for instance done for quark field correlators in Ref.~\cite{bm98}, one
anyway reduces the matrix elements to lightlike separations.

We note that in different processes, e.g. lepton-hadron scattering or
Drell-Yan scattering the paths will run from different points,
that is $\xi^-$ = $\infty$ and $-\infty$, respectively. In $A^+ = 0$
gauge, the difference,
however, is precisely $A^i_{\scriptscriptstyle T}(\xi^- = \infty) +
A_{\scriptscriptstyle T}^i(\xi^- = -\infty)$, a quantity remaining to be
fixed to fully fix the lightcone gauge~\cite{kogut70,bmt}.


\section{The Twist Two Functions as densities} \label{twisttwo}

The fact that we in the gauge $A^+ = 0$  are left with $S^{ij}$, 
a matrix element bilinear
in the gluonic fields, suggests that it might be possible
to find a probabilistic interpretation for some distribution functions. 
This has been discussed in detail in several papers, expanding
the transverse gluon fields in modes. We follow here a slightly
different route that allows us to draw conclusions on the newly
introduced leading twist transverse momentum dependent functions 
of the previous section.

The basic idea is the observation that 
$M^{ij} \equiv (2M/x \pp) \Gamm^{ij}$ is a two by two matrix
in the two transverse polarizations that for any diagonal element
is a (positive-definite) density. Generalizing also to a matrix
in the hadron spin space, one has 

\bea
\Gamma^{ij}_{\Lambda \Lambda^\prime}(x) & = &
\left. \int \frac{d\xi^-}{2\pi}\ e^{ik\cdot \xi} \langle P,\Lambda \vert
F^{+i}(0)\,F^{+j}(\xi)\vert P,\Lambda^\prime\rangle\right|_{\xi^+
=\xi_{\scriptscriptstyle T}=0} \nn \\
& = &
\sum_n \langle P_n\vert F^{+i}(0)\vert P,\Lambda\rangle^*
\langle P_n\vert F^{+j}(0)\vert P,\Lambda^\prime\rangle
\,\delta\left(P_n^+ - (1-x)P^+\right).
\eea

In principle it does not matter for our considerations if we use the
matrix for (real) linear polarizations or for the circular polarizations 
of the gluons. For interpretational purposes, the latter however is
more common. Using the circular polarizations

\be
\vert \pm\rangle = \mp \frac{1}{\sqrt{2}}\left(\vert x\rangle
\pm i\,\vert y\rangle\right),
\ee

\nin we obtain the matrix elements

\bea
M^{++} & = & \half (M^{11} + M^{22}) - {\cal I}m \,M^{12}, \nn \\
M^{+-} & = & - \half (M^{11} - M^{22}) + i\,{\cal R}e \,M^{12}, \nn \\
M^{-+} & = & - \half (M^{11} - M^{22}) - i\,{\cal R}e \,M^{12}, \nn \\
M^{--} & = & \half (M^{11} + M^{22}) + {\cal I}m \,M^{12}. \nn 
\eea

Explicitly we find from the parameterization in Eqs~\ref{par1} - \ref{par3}
the matrix elements (in gluon polarization space), 
\bea
M^{++} & = & G + \sL \ \Del G_L 
- S_T^1 \ {|\ktns| \over M} (\sin \phi \ G_T - \cos \phi \ \Del G_T)
+ S_T^2 \ {|\ktns| \over M} (\cos \phi \ G_T + \sin \phi \ \Del G_T), \\
M^{+-} & = & - {\kts \over 2 M^2} \ e^{-2i \phi} H^\perp 
-i \sL \ {\kts \over 2 M^2} \ e^{-2i \phi} \Del H_L^\perp 
- i (S_T^1+i\,S_T^2) {\kts \over 4 M^2} \ e^{-3i \phi} {|\ktns| \over M} 
\ \Del H_T^\perp
\nn \\
& & - i e^{-i \phi} \ {|\ktns| \over 2M} \ \Del H_T 
(S_T^1-i\,S_T^2)
, \\
M^{-+} & = & - {\kts \over 2 M^2} \ e^{+2i \phi} H^\perp 
+i \sL \ {\kts \over 2 M^2} \ e^{+2i \phi} \Del H_L^\perp 
+ i (S_T^1-i\,S_T^2) {\kts \over 4 M^2} \ e^{+3i \phi} {|\ktns| \over M} 
\ \Del H_T^\perp
\nn \\
& & + i e^{+i \phi} 
\ {|\ktns| \over 2M} \ \Del H_T (S_T^1+i\,S_T^2), \\
M^{--} & = & G - \sL \ \Del G_L 
- S_T^1 \ {|\ktns| \over M} (\sin \phi \ G_T + \cos \phi \ \Del G_T)
+ S_T^2 \ {|\ktns| \over M} (\cos \phi \ G_T - \sin \phi \ \Del G_T).
\eea

In order to make the nucleon spin explicit we use the connection

\be \label{elena1}
\Gamm^{ij}\xktnsps = \sum_{\Lam,\Lam'} \rho_{\Lam' \Lam} (\bfS) \ 
\Gamm^{ij}_{\Lam \Lam'} \xktnsps,
\ee

\nin where $\rho (\bfS)$ is the spin density matrix for a spin $1/2$
particle characterized by the spin vector $\bfS=(\sL, \bfS_T)$, 
in its rest frame given by

\be \label{density}
\rho (\bfS) = \half ({\bf 1} + \bfS \cdot \bm \sig).
\ee

Using the explicit form of the density matrix we can write (\ref{elena1}) as 

\be \label{elena2}
\Gamm^{ij} = \half \left( \Gamm^{ij}_{++} + \Gamm^{ij}_{--} \right)
+ {\sL \over 2} \left( \Gamm^{ij}_{++} - \Gamm^{ij}_{--} \right)
+ {S_T^1 \over 2} \left( \Gamm^{ij}_{+-} + \Gamm^{ij}_{-+} \right)
+ {i S_T^2 \over 2} \left( \Gamm^{ij}_{+-} - \Gamm^{ij}_{-+} \right).
\ee

We shall now apply this to the matrix $M^{ij}$, 
where we include also T-odd functions. 
We then obtain a $4 \times 4$ matrix in the gluon $\otimes$ nucleon
spin space [$\vert gluon;nucleon\rangle$ basis $\vert +;+\rangle$,
$\vert +;-\rangle$, $\vert -;+\rangle$ and $\vert -;-\rangle$],
\bea 
&&
\left(
\ba{cccc}
G + \Del G_L & 
{|\ktns| e^{-i \phi} \over M} \left[\Del G_T -i G_T  \right] & 
-e^{-2i \phi} \left[ H^{\perp (1)} + i\Del H_L^{\perp (1)} \right] & 
-i {|\ktns| e^{-3i \phi} \over M} \ \Del H_T^{\perp (1)}  \\
{|\ktns| e^{i \phi} \over M} \left[ \Del G_T + i G_T \right] & 
G - \Del G_L & 
-i {|\ktns| e^{-i \phi} \over M} \ \Del H_T & 
- e^{-2i \phi} \left[ H^{\perp (1)} - i \Del H_L^{\perp (1)} \right] \\
-e^{2i \phi} \left[ H^{\perp (1)} - i\Del H_L^{\perp (1)} \right] & 
i {|\ktns| e^{i \phi} \over M} \ \Del H_T &
G - \Del G_L & 
-{|\ktns| e^{-i \phi} \over M} \left[ \Del G_T + i G_T \right] \\
i {|\ktns| e^{3i \phi} \over M} \ \Del H_T^{\perp (1)}  &
-e^{2i \phi} \left[ H^{\perp (1)} + i \Del H_L^{\perp (1)} \right] &
-{|\ktns| e^{i \phi} \over M} \left[ \Del G_T - i G_T \right] &
G + \Del G_L
\ea
\right)
\nn \\ &&
\label{matrix}
\eea

The matrix representation is also convenient to find the physical meaning
of the distributions.
Well known is $G$ which measures the number of gluons with momentum $(x,
\ktns)$ in a hadron. The functions $\Del G_L$ ($\Del G_T$) represents
the difference of the numbers of gluons with opposite circular
polarizations in a longitudinally (transversely) polarized nucleon.
The off-diagonal function $H^\perp$ also is a difference of
densities, but in this case of linearly polarized gluons in an unpolarized
hadron. Using the circular polarizations, $H^\perp$ flips the polarization.


\section{Gluon Fragmentation Functions} \label{frags}

\begin{figure}[t]
\begin{center}
\epsfig{file=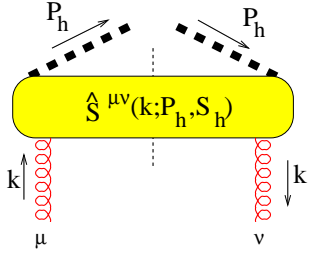,width = 4.0 cm}
\end{center}
\caption{\label{fig3}
The gluon correlator}
\end{figure}

The procedure to analyse the gluon fragmentation functions is
very similar to what has been done for the distributions in the previous
sections.

Using as parameterization of the vectors

\bea
P_h & = & P_h^- \nm + {M_h^2 \over 2P_h^-} \np, \\
k & = & {P_h^- \over z} \nm + {z (k^2 + \kts) \over 2 P_h^-} \np + \ktnsnb, \\
S_h & = & \shL {P_h^- \over M_h} \nm 
- \shL {M_h \over 2 P_h^-} \np + S_{hT}.
\eea

\nin
for a gluon with momentum $k$ fragmenting into a hadron with momentum
$P_h$ and spin $S_h$, we consider the soft part (see Fig.~\ref{fig3})

\be
\hat S^{\mu\nu}(k;P_h,S_h)
= \sum_X \int \frac{d^4\xi}{(2\pi)^4}\,e^{ik\cdot \xi}
\,\langle 0\vert A^\nu (\xi) \vert P_h,S_h; X\rangle 
\langle P_h,S_h;X\vert A^\mu(0)\vert 0\rangle,
\ee

\nin
or the appropriate gauge-invariant object

\be
\hat \Gamma^{\mu\nu;\rho\sigma}(k;P_h,S_h)
= \sum_X \int \frac{d^4\xi}{(2\pi)^4}\,e^{ik\cdot \xi}
\,\langle 0\vert  F^{\rho\sigma}(\xi) \vert P_h,S_h; X\rangle 
\langle P_h,S_h;X\vert {\cal U}(\xi,0)\,F^{\mu\nu}(0)\vert 0\rangle .
\ee

For the description of fragmentation in leading order in the inverse
hard scale, we need this correlation function integrated over one
lightcone direction, with the above choice for $P_h$, being the momentum
$k^+$,
\be
M_h\hat \Gamm^{ij}(z,\bm k_\subt) =
\int dk^+ \ \Gamma^{-j;-i}(k;P_h,S_h),
\ee
which, separating the polarizations, is parameterized as
\bea
\label{frag1}
\hat \Gamm_O^{ij} \zktns & = &  {P_h^- \over M_h} 
\left[ - \gtij \hat G \zkp 
+ \left( {\ktnsnb^i \ktnsnb^j \over M_h^2} + \gtij {\kts \over 2M_h^2} \right)
\hat H^\perp \zkp \right], \\
\label{frag2}
\hat \Gamm_L^{ij} \zktns & = & {P_h^- \over M_h} 
\left[ i \epstij \shL \ \Del \hat G_L \zkp
- {\epst^{\ktnsnb \left\{ i \right. } \ktnsnb^{\left. j \right\}}\over 2M_h^2}
\ \shL \ \Del \hat H_L^\perp \zkp \right], \\
\label{frag3}
\hat \Gamm_T^{ij} \zktns & = & {P_h^- \over M_h} 
\Biggl[ \gtij \ {\epst^{\ktnsnb \shtnsnb} \over M_h} \ \hat G_T \zkp
+ i \epstij \ {\ktns \cdot \shtns \over M_h} \ \Del \hat G_T \zkp \nn \\
& & \qquad\quad 
- {\epst^{\ktnsnb \{ i } \ktnsnb^{j \}} \over 2M_h^2}
{\ktns \cdot \shtns \over M_h} \ \Del \hat H_T^\perp \zkp 
\nn \\ & & \qquad\quad 
- {\epst^{\ktnsnb \left\{ i \right. } \shtnsnb^{\left. j \right\}} +
\epst^{\shtnsnb \left\{ i \right. } \ktnsnb^{\left. j \right\}} \over 4M_h}
\,\left[ \Del \hat H_T \zkp - \frac{\kts}{2M^2}\,\Del \hat H_T^\perp
\zkp \right] \Biggr],
\eea

\nin
where the argument of the fragmentation functions, $\bm k_\subt^\prime
= - z\bm k_\subt$ is the transverse momentum of the hadron with respect
to the gluon. The factors are chosen such that $\int
dz\,d^2k_{\scriptscriptstyle T}^\prime\ z\hat G \zkp = \langle z\rangle_h$
is the fraction of momentum of the struck gluon taken by the hadron $h$,
for which we have $\sum_h \langle z\rangle_h = 1$.
While for distribution functions T-odd functions might appear via special
mechanisms dealing with initial state interactions or gluonic poles, 
this is not the case for fragmentation functions~\cite{hhk83,Jaffe-Ji-93}, 
where time-reversal symmetry
cannot be used as a constraint because of the explicit appearance of out
states $\vert P_h; X\rangle$ in the definition. Thus one expects nonvanishing
fragmentation functions $\hat G_T \zkp$, $\Delta \hat H_L^\perp \zkp$, $\Delta
\hat H_T^\perp \zkp$ and $\Delta \hat H_T \zkp$.


\section{Bounds on the distribution and fragmentation functions}

As discussed before, we can organize the distribution functions in a
matrix representation in the gluon $\otimes$ nucleon spin space. For
distribution functions one has (omitting the T-odd functions) a matrix

\be 
\frac{2}{x}\,\frac{M}{P^+}\,\Gamma (x,\ktns) =
\left(
\ba{cccc}
G + \Del G_L & 
{|\ktns| e^{-i \phi} \over M} \ \Del G_T  & 
-e^{-2i \phi} \  H^{\perp (1)}  & 
0 \\
{|\ktns| e^{i \phi} \over M} \ \Del G_T  & 
G - \Del G_L & 
0 &
- e^{-2i \phi} \ H^{\perp (1)}  \\
-e^{2i \phi} \ H^{\perp (1)}  & 
0 &
G - \Del G_L & 
-{|\ktns| e^{-i \phi} \over M} \  \Del G_T  \\
0 &
-e^{2i \phi} \ H^{\perp (1)}  &
-{|\ktns| e^{i \phi} \over M} \ \Del G_T  &
G + \Del G_L
\ea
\right).
\ee

\nin
Requiring any diagonal element to be positive gives using the diagonal
elements the trivial bound
\be
\vert \Delta G_L \xkts \vert \le G \xkts .
\ee
Using all possible 2 $\times$ 2 submatrices, positivity leads to bounds
\bea
&&
\vert \Delta G_T^{(1)} \vert \le 
\frac{\vert \ktns \vert}{2M}\,\sqrt{(G + \Delta G_L)(G - \Delta G_L)}
\le 
\frac{\vert \ktns \vert}{2M}\,G ,
\\ &&
\vert H^{\perp (1)} \vert \le 
\sqrt{(G + \Delta G_L)(G - \Delta G_L)} \le G .
\eea
These bounds can still be sharpened by using the eigenvalues of the full
4 $\times$ 4 matrix in analogy to what was done for quark distributions
in Ref.~\cite{bacchetta00}.

For fragmentation functions the matrix 
$(M_h/P_h^-)\,\hat \Gamma(z,\bm k_\subt)$
contains the various fragmentation functions, now including the T-odd
functions. It thus is the same as the matrix in \ref{matrix}, but with hat
functions depending on $z$ and $\bm k_\subt^\prime$. The constraints become
\bea
&&\vert \Delta \hat G_L \zkp \vert \le \hat G \zkp ,
\\ &&
\vert \Delta \hat G_T^{(1)} \vert^2 + 
\vert \hat G_T^{(1)}\vert^2 \le 
\frac{\vert k_\subt\vert^2}{4M_h^2}\,(\hat G + \Delta \hat G_L)(\hat G - \Delta
\hat G_L) \le 
\frac{\vert k_\subt\vert^2}{4M_h^2}\,\hat G^2 ,
\\ &&
\vert \hat H^{\perp (1)} \vert^2 
+ \vert \Delta \hat H_L^{\perp (1)}\vert^2 \le 
(\hat G + \Delta \hat G_L)(\hat G - \Delta \hat G_L) \le \hat G^2 ,
\\ &&
\vert \Delta \hat H_T^{\perp (2)} \vert \le
\frac{\vert \ktns \vert}{2M_h}\,(\hat G + \Delta \hat G_L) ,
\\ &&
\vert \Delta \hat H_T^{(1)}\vert \le 
\frac{\vert \ktns \vert}{2M_h}\,(\hat G - \Delta \hat G_L) .
\eea


\section{Conclusions}

\begin{figure}[t]
\begin{center}
\begin{minipage}{5.0cm}
\begin{center}
\epsfig{file=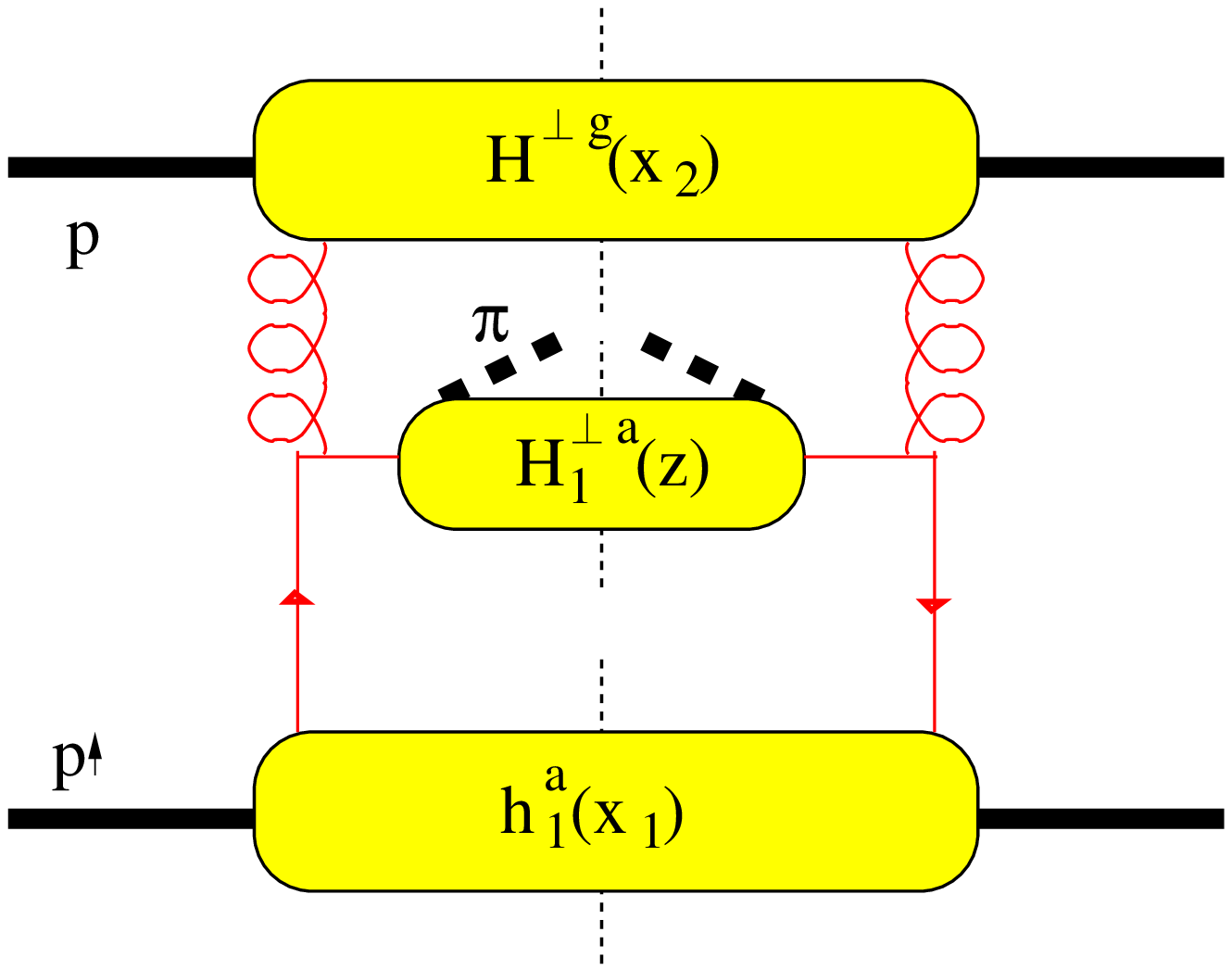,width = 4.5 cm}
\\
(a)
\end{center}
\end{minipage}
\hspace{0.5cm}
\begin{minipage}{5.0cm}
\begin{center}
\epsfig{file=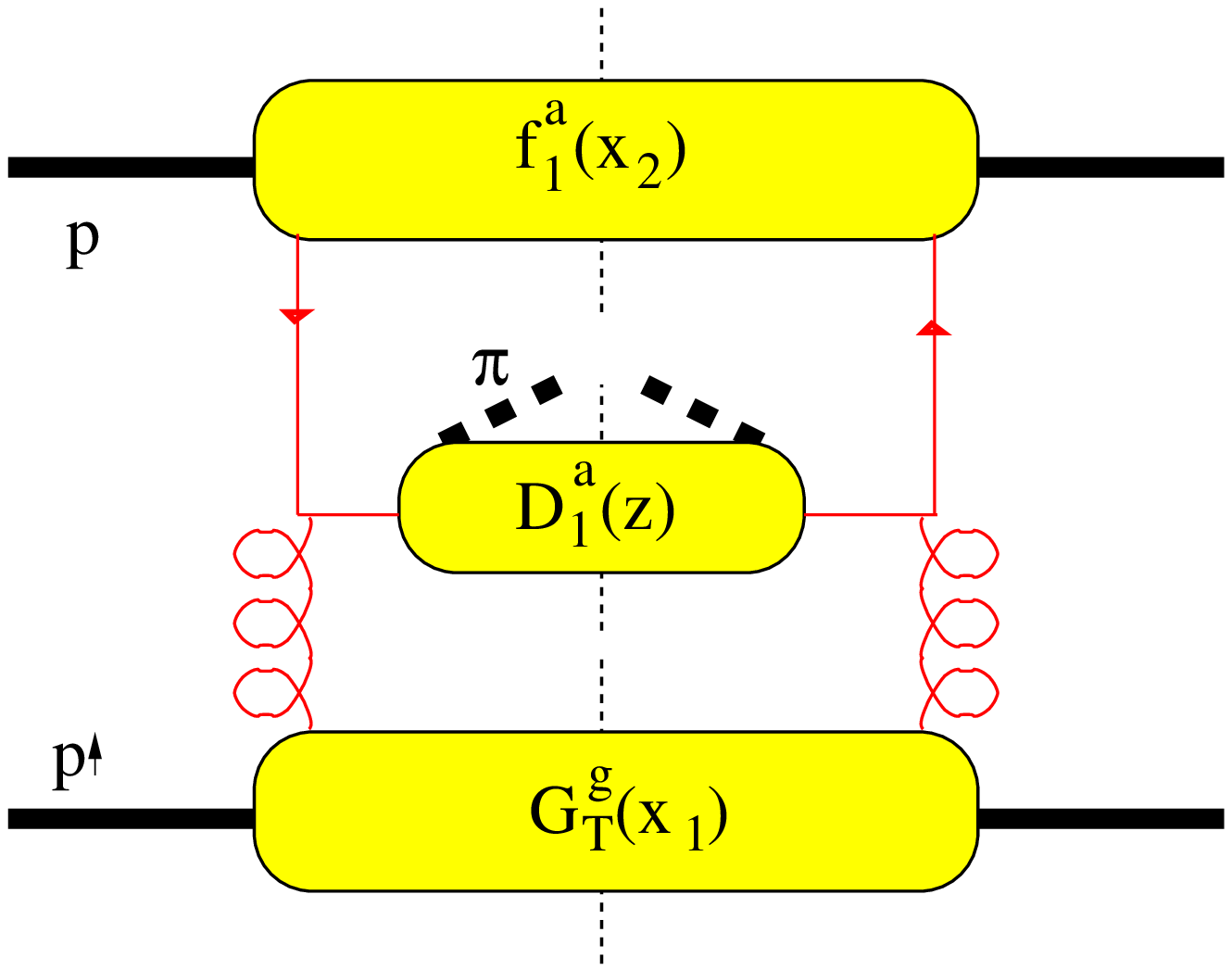,width = 4.5 cm}
\\
(b)
\end{center}
\end{minipage}
\hspace{0.5cm}
\begin{minipage}{5.0cm}
\begin{center}
\epsfig{file=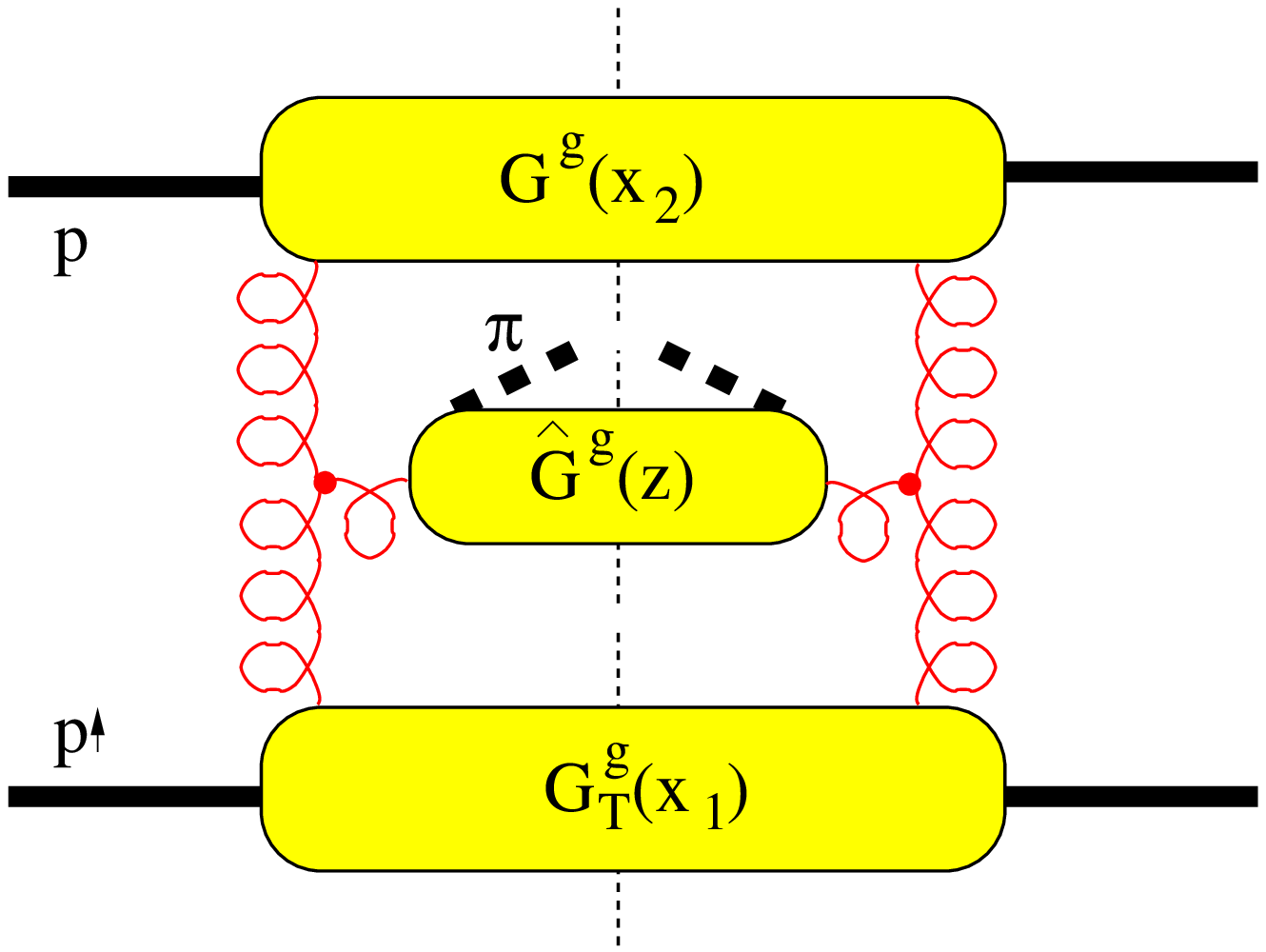,width = 4.5 cm}
\\
(c)
\end{center}
\end{minipage}
\end{center}
\caption{\label{fig4}
Gluon distribution and fragmentation functions contributing to the
single spin asymmetry in $p^\uparrow p \rightarrow \pi X$ scattering
cross section.}
\end{figure}

We have given a full classification of the gluon distributions and 
fragmentation functions relevant in
hard scattering processes in leading order in the inverse 
hard scale including transverse momentum dependence. Some results
for subleading (twist three) correlation functions have been given
also. The inclusion of transverse momentum dependence is
needed in processes involving at least two hadrons. Examples of such
processes are 1-particle inclusive leptoproduction or Drell-Yan scattering.
In these processes one can become sensitive to transverse momentum dependence,
in particular when one considers azimuthal dependence in the final 
state~\cite{piet95,piet96,bm98,bjm00}. We note that in electroweak processes,
the gluon correlation functions do not enter at tree level but only at higher
order in $\alpha_s$. This also implies their relevance in the study of
evolution of the transverse momentum dependent flavor-singlet quark
distribution and  fragmentation functions. Also in related
processes, e.g. $\chi_2$-production in hadron-hadron scattering the relevance
of gluon distribution functions has been emphasized and
investigated~\cite{hoodbhoy93}. Since we have discussed both distribution and
fragmentation functions we have also classified the T-odd functions, important
in the latter case. In single spin asymmetries at least one of the functions
describing the soft physics is a T-odd function.

In order to illustrate the importance of also considering gluons, we
consider the contributions to pion production in $pp^\uparrow$ scattering
in which one of the protons is
polarized~\cite{adams,am98,abm99,sivers,boglione00}. 
In this process a large single spin asymmetry is found. 
We note that gluon correlation functions can play an important role here. 
This is another way of looking to the approach in Refs.
\cite{am98,abm99,sivers,boglione00}.
Since in this case several mechanisms are considered in these papers,
we need both T-odd and T-even distribution and fragmentation functions.
In Fig.~\ref{fig4} three contributions are shown
of gluon correlation functions producing an asymmetry. 
Actually kinematics
requires one of the three partons to be off-shell, which means that we need
the asymptotic transverse momentum dependence of the soft parts, i.e. the
evolution of the soft parts. Nevertheless the structure of the gluonic soft
parts is sufficient to indicate that diagram (4a) will produce a  $\sin
(\phi_\pi + \phi_S)$ asymmetry (being equivalent to the Collins asymmetry in
leptoproduction). This asymmetry is proportional to $(h_1^a (x_1)\times
H^\perp(x_2))\times \hat h_1^{\perp}(z)$ involving the transverse momentum
dependent gluon distribution function $H^\perp$, the transverse spin
distribution $h_1^a$ and the Collins function $\hat h_1^\perp$ (we use here
the hat-notation for the fragmentation function in order to avoid
confusion between fragmentation functions and gluon distribution functions).
Diagram (4b) will lead to a $\sin (\phi_\pi - \phi_S)$ asymmetry proportional
to $(G_T(x_1) \times f_1^a (x_2))\times D_1^a(z)$ involving a T-odd
gluon distribution function, the unpolarized quark distribution function
$f_1^a$ and fragmentation function $D_1^a$. Diagram (4c) gives a
similar asymmetry as (4b) with the unpolarized gluon distribution function
$G$ and gluon fragmentation function $\hat G$.


\acknowledgments

This work is supported by the Foundation for Fundamental Research on Matter
(FOM), the National Organization for Scientific Research (NWO) and the Junta
Nacional de Investiga\c{c}\~{a}o Cient\'{\i}fica (JNICT, PRAXIS XXI).


\appendix

\section{Gluon Distributions} \label{appex1}

Leading and subleading gluon correlations are distinguished via the
Lorentz indices. It is therefore convenient to rewrite the covariant
expression in Eq.~\ref{amplitudes}
in terms of $(\np, \nm, \ktns, \stns)$. We start with the correlation
functions with the maximal (that is two) number of plus indices.
Distinguishing
unpolarized (O), longitudinally polarized (L) and transversely 
polarized (T) situations (spin 1/2),

\be
\Gamm^{+\mu ; +\nu} = \Gamm_O^{+\mu ; +\nu} + 
\Gamm_L^{+ \mu ; + \nu} + \Gamm_T^{+ \mu ; + \nu} ,
\ee

\nin one has, with the (dimensionless) invariants 
$\sig = 2k\cdot P/M^2$ and $\tau=k^2/M^2$, the result
\bea
\label{app1}
\Gamm_O^{+ \mu ; + \nu} & = & 
\left( {\pp \over M} \right)^2 (- \gtmn) \left[ (X_2 + x X_4) 
+ x ( X_4 + x X_3 ) \right]
+ \left( {\pp \over M} \right)^2 {\ktnsnbmu \ktnsnbnu \over M^2} \ X_6 \nn \\
& & + {\pp \over M} \ 
{\ktnsnb^{ \left\{ \mu \right. } \nm^{ \left. \nu \right\} } \over M}
\left[ (X_4 + x X_3) + \left( {\sig \over 2} - x \right) X_6  \right]
- {\pp \over M} \ 
{\ktnsnb^{ \left[ \mu \right. } \nm^{ \left. \nu \right] } \over M}
\ i X_5 \nn \\
& & + \nm^\mu \nm^\nu \left[ 2 X_1 
+ [(X_2 + x X_4) + x (X_4 + x X_3)] 
+ 2 \left( {\sig \over 2} - x \right) (X_4 + x X_3 )
+ \left( {\sig \over 2} - x \right)^2 X_6 \right], \\
\Gamm_L^{+ \mu ; + \nu} & = & i \sL \left( \ppoverm \right)^2 \epstmn 
\left\{ (X_8 + x X_{10}) + (X_9 + x X_{11}) 
+ \left( {\sig \over 2} - x \right) 
\left[ (X_{12} + x X_{14}) + x (X_{15} + x X_{13}) \right] \right. \nn \\
& & \left. - {\kts \over M^2} 
\left[ X_{19} + X_{23} + \left( {\sig \over 2} - x \right) X_{25} \right]
\right\} \nn \\
& & + \sL \left( \ppoverm \right)^2
{\epst^{ \ktnsnb \left\{ \mu \right. } \ktnsnb^{ \left. \nu \right\}}
\over M^2} \left[ - (X_{18} + X_{22}) 
- \left( {\sig \over 2} - x \right) X_{24} \right] \nn \\
& & + i \sL \ \ppoverm \ 
{\eps^{\ktnsnb \left[ \mu \right.} \nm^{\left. \nu \right]} \over M}
\left[ X_{10} + \left( {\sig \over 2} - x \right) (X_{15} + x X_{13}) 
\right. \nn \\
& & \left. - \left( {\sig \over 2} - x \right) (X_{19} + X_{23})
- (X_{21} + x X_{23}) - \left( {\sig \over 2} - x \right)^2 X_{25} \right]
\nn \\
& & + \sL \ \ppoverm
{\epst^{ \ktnsnb \left\{ \mu \right. } \nm^{ \left. \nu \right\}} \over M} 
\left[ 
- \left( {\sig \over 2} - x \right) (X_{18} + X_{22}) - (X_{20} + x X_{22} )
- \left( {\sig \over 2} - x \right)^2 X_{24} \right], \\
\Gamm_T^{+ \mu ; + \nu} & = & 
i \left( \ppoverm \right)^2 \ktstoverm \ \epstmn \left[
- (X_{12} + x X_{14}) - x (X_{15} + x X_{13}) \right. \nn \\
& & \left. + (X_{17} + x X_{19}) + (X_{21} + x X_{23}) 
+ {\kts \over M^2} \ X_{25} \right] \nn \\
& & + \left( {\pp \over M} \right)^2 \left\{ 
{\eps_T^{\ktnsnb \left\{ \mu \right. } \ktnsnb^{ \left. \nu \right\} }
\over M^2}
\ {\ktns \cdot \stns \over M} \ X_{24} + 
{\eps_T^{\ktnsnb \left\{ \mu \right. } \stnsnb^{ \left. \nu \right\}} \over M}
\left( X_{20} + x X_{22} \right) + 
{\eps_T^{\stnsnb \left\{ \mu \right. } \ktnsnb^{ \left. \nu \right\} }\over M}
\left( X_{16} + x X_{18} \right) \right\} \nn \\
& & + i \ppoverm \left\{
{ \epst^{ \ktnsnb \left[ \mu \right. } \nm^{ \left. \nu \right]} \over M}
\ \ktstoverm \left[ -(X_{15} + x X_{13}) + X_{19} +
\left( {\sig \over 2} - x \right) X_{25} \right] \right. \nn \\
& & \left. + \epst^{ \stnsnb \left[ \mu \right. } \nm^{ \left. \nu \right]}
\left[ (X_9 + x X_{11}) + \left( {\sig \over 2} - x \right) (X_{17} + x X_{19})
- {\kts \over M^2} \ X_{19} \right] \right\} \nn \\
& & + {\pp \over M} \left\{ 
{\eps_T^{\ktnsnb \left\{ \mu \right. } \nm^{ \left. \nu \right\} } \over M}
\ {\ktns \cdot \stns \over M} \left( {\sig \over 2} - x \right) X_{24}
+ \epst^{\stnsnb \left\{ \mu \right. } \nm^{ \left. \nu \right\} }
\left( {\sig \over 2} - x \right) (X_{16} + x X_{18})
- {\eps^{\nm \ktnsnb \stnsnb \left\{ \mu \right. } \ktnsnb^{\left. \nu
\right\} } \over M^2} \ X_{18} \right\} \nn \\
& & - {\eps^{\nm \ktnsnb \stnsnb \left\{ \mu \right. } \nm^{\left. \nu
\right\} } \over M} \left( {\sig \over 2} - x \right) X_{18},
\eea

We need the soft parts integrated over the momentum component $k^-$.  
Upon integration over $dk^-$ we find that the functions appearing in the 
decomposition of the quantity $M\Gamm^{ij}$ in Eq.~\ref{l22} can be 
expressed in the amplitudes of Eq.~\ref{amplitudes}. For this we use

\be 
M \Gamm^{ij}(x, \ktns;P,S) \equiv \int dk^- \ 
\Gamm^{+ i; + j} (k;P,S) 
=
{M^2 \over 2 \pp} \ \dsdt \ \Gamm^{+ i; + j} (k;P,S),
\ee

\nin where we have introduced the shorthand notation

\be
[d \sig d\tau ] \equiv d \sig d\tau \ 
\del \left( \tau - x \sig + x^2 + {\kts \over M^2} \right),
\ee

\nin to indicate integration over $\sig$ and $\tau$. The results for
the twist two distributions in terms of the amplitudes $X_i$ are the
following: 
\bea
x G \xkts & = & \dsdt \ 
\left[ [(X_2 + x X_4) + x (X_4 + x X_3)]
+ {\kts \over 2 M^2} \ X_6 \right], \nn \\
x G_T \xkts & = & \dsdt \ \left[ (X_{16} + x X_{18}) - 
(X_{20} + x X_{22}) \right], \nn \\
x H^\perp \xkts & = & \dsdt \ X_6, \nn \\
x \Del G_L \xkts & = & -\dsdt \ \left[ 
(X_8 + x X_{10}) + (X_9 + x X_{11}) 
+ \left( {\sig \over 2} - x \right) 
[(X_{12} + x X_{14}) + x (X_{15} + x X_{13})] \right. \nn \\
& & \left. - {\kts \over M^2} \left[ X_{19} + X_{23} 
+ \left( {\sig \over 2} - x \right) X_{25} \right] \right], \nn \\
x \Del G_T \xkts & = &  \dsdt \ \left[ 
(X_{12} + x X_{14}) + x (X_{15} + x X_{13})
- (X_{17} + x X_{19}) - (X_{21} + x X_{23} )
- {\kts \over M^2} \ X_{25} \right], \nn \\
x \ \Del H_L^\perp \xkts & = & -2 \dsdt \ \left[
(X_{18} + X_{22}) 
+ \left({\sig \over 2} - x \right) X_{24} \right], \nn \\
x \Del H_T^\perp \xkts & = & 2 \dsdt \ X_{24}, \nn \\
x \Del H_T \xkts & = & 2 \dsdt \ \left[ (X_{16} + x X_{18}) + 
(X_{20} + x X_{22}) - {\kts\over 2M^2}\ X_{24}\right]. \nn 
\eea

{}From Eq.~\ref{app1} we can also obtain the relation between
the functions in 
the first quantity, $M\Gamm^{i-}$, relevant at twist three.
We find that all functions have a real part (which
is T-even) and an imaginary part (which is T-odd):
\bea
{\cal R}e \left( x G_3^\perp \xkts \right) & = & 
\dsdt \ \left[ (X_4 + x X_3)
+ \left({\sig \over 2}- x \right) X_6 \right], \nn \\
{\cal I}m \left( x G_3^{\perp} \xkts \right) & = & 
\dsdt \ \left[ -X_5 \right] , \nn \\
{\cal R}e \left( x \Del G_{3L}^\perp \xkts \right) & = & \dsdt \ \left[ 
X_{10} + \left( {\sig \over 2} - x \right) (X_{15} + x X_{13}) 
- \left( {\sig \over 2} - x \right) (X_{19} + X_{23}) \right. \nn \\
& & \left. 
- (X_{21} + x X_{23}) - \left( {\sig \over 2} - x \right)^2 X_{25} \right], 
\nn \\
{\cal I}m \left( x \Del G_{3L}^{\perp} \xkts \right) & = & \dsdt \ \left[
\left( {\sig \over 2} - x \right) (X_{18} + X_{22}) + (X_{20} + x X_{22} )
+ \left( {\sig \over 2} - x \right)^2 X_{24} \right], \nn \\
{\cal R}e \left( x \Del G_{3T} \xkts \right) & = & \dsdt \ 
\left[ (X_9 + x X_{11}) + \left( {\sig \over 2} - x \right) (X_{17} + x X_{19})
\right. \nn \\
& & \left. + {\kts \over 2M^2} \left[ - (X_{15} + x X_{13})
- X_{19} + \left( {\sig \over 2} - x \right) X_{25} \right] \right], \nn \\
{\cal I}m \left( x \Del G_{3T} \xkts \right) & = & \dsdt \ 
\left[ - \left( {\sig \over 2} - x \right) (X_{16} + x X_{18})
- {\kts \over 2 M^2} \left[ X_{18} + 
\left( {\sig \over 2} - x \right) X_{24} \right] \right], \nn \\
{\cal R}e \left( x \Del G_{3T}^\perp \xkts \right) 
& = & \dsdt \ \left[
- (X_{15} + x X_{13}) + X_{19} 
+ \left( {\sig \over 2} - x \right) X_{25} \right] \nn \\
{\cal I}m \left( x \Del G_{3T}^{\perp} \xkts \right) & = & \dsdt 
\ \left[ X_{18} 
- \left( {\sig \over 2} - x \right) X_{24} \right]. \nn
\eea


We now turn our attention to the  different set of indices, namely
$(ij;l+)$, and the corresponding correlator $\Gamm^{ij,l+}(k;P,S)$. 
We again decompose according to the spin of the hadron and find
\bea
\Gamm_O^{ij;l+} & = & \ppoverm \left\{
{- \gt^{l \left[ i \right. } \ktnsnb^{\left. j \right] } \over M}
\left[ (X_4 + x X_3) - i X_5 \right] \right\}, \nn \\
\Gamm_L^{ij;l+} & = & \ppoverm \left\{
i \sL \epstij \ {\ktnsnb^l \over M} \left[ 
X_{11} + \left( {\sig \over 2} - x \right) (X_{14} + x X_{13})
- (X_{17} + x X_{19}) \right. \right. \nn \\
& & \left. \left. 
- \left( {\sig \over 2} - x \right) (X_{19} + X_{23})
- \left( {\sig \over 2} - x \right)^2 X_{25} \right] \right. \nn \\
& & \left. 
+ \sL \epstij \ {\ktnsnb^l \over M} \left[ 
- (X_{16} + x X_{18})
- \left( {\sig \over 2} - x \right) (X_{18} + X_{22})
- \left( {\sig \over 2} - x \right)^2 X_{24} \right] \right\}, \\
\Gamm_T^{ij;l+} & = & \ppoverm \left\{
i \epstij \ {\ktnsnb^l \over M} \ \ktstoverm
\left[ -(X_{14} + x X_{13}) + X_{23} 
+ \left( {\sig \over 2} - x \right) X_{25} \right]
+ \epstij \ {\ktnsnb^l \over M} \ \ktstoverm
\left[ -X_{22} +\left( {\sig \over 2} - x \right) X_{24} \right] \right. \nn \\
& & \left. + i \epstij \stnsnb^l \left[ (X_8 + x X_{10}) + 
\left( {\sig \over 2} - x \right) (X_{21} + x X_{23}) 
- {\kts \over M^2} \ X_{23} \right] 
+ \epstij \stnsnb^l  \left[
\left( {\sig \over 2} - x \right) (X_{20} + x X_{22}) 
+ {\kts \over M^2} \ X_{22} \right] \right\} \nn .
\eea

Upon integration over $dk^-$, we find that the functions appearing in

\be
M \Gamm^{ij,l} (x,\ktns) \equiv \int dk^- \ \Gamm^{ij,l+}(k;P,S) = 
{M^2 \over 2 \pp} \ \dsdt \ \Gamm^{ij,l+}(k;P,S)
\ee

\nin can be expressed in the amplitudes as follows:

\bea
{\cal R}e \left( x H_3^\perp \xkts \right) 
& = & \dsdt \ \left[ X_4 + x X_3 \right], \nn \\
{\cal I}m \left( x H_3^\perp \xkts \right) 
& = & \dsdt \ \left[ -X_5 \right], \nn \\
{\cal R}e \left( x \Del H_{3L}^\perp \xkts \right) 
& = & \dsdt \ \left[ X_{11} + \left( {\sig \over 2} - x \right) 
(X_{14} + x X_{13})
- (X_{17} + x X_{19}) \nn \right. \\
& & \left.  - \left( {\sig \over 2} - x \right) (X_{19} + X_{23})
- \left( {\sig \over 2} - x \right)^2 X_{25} \right], \nn \\
{\cal I}m \left( x \Del H_{3L}^\perp \xkts \right) 
& = & \dsdt \ \left[ (X_{16} + x X_{18})
+ \left( {\sig \over 2} - x \right) (X_{18} + X_{22})
+ \left( {\sig \over 2} - x \right)^2 X_{24} \right], \nn \\
{\cal R}e \left( x \Del H_{3T}^\perp \xkts \right) & = & 
\dsdt \ \left[ 
-(X_{14} + x X_{13}) + X_{23}
+ \left( {\sig \over 2} - x \right) X_{25} \right], \nn \\
{\cal I}m \left( x \Del H_{3T}^\perp \xkts \right) 
& = & \dsdt \ \left[ X_{22} -
\left( {\sig \over 2} - x \right) X_{24} \right], \nn \\
{\cal R}e \left( x \Del H_{3T} \xkts \right) 
& = & \dsdt \ \left\{ (X_8 + x X_{10}) + 
\left( {\sig \over 2} - x \right) (X_{21} + x X_{23}) \right. \nn \\
& & \left. + {\kts \over 2M^2} \left[ -(X_{14}+xX_{13}) - X_{23} 
+ \left( {\sig \over 2} - x \right) X_{25} \right] \right\}, \nn \\
{\cal I}m \left( x \Del H_{3T} \xkts \right) 
& = & \dsdt \ \left\{
- \left( {\sig \over 2} - x \right) (X_{20} + x X_{22}) 
- {\kts \over 2M^2} \left[ X_{22} + \left( {\sig \over 2} - x \right) X_{24}
\right] \right\}.
\eea

The integrated functions in terms of the amplitudes are given by
\bea
x G(x) & = & M^2 \pi \dsdtau 
\ \left[ (X_2 + xX_4) + x (X_4 + x X_3)
+ ( x \sig - x^2 - \tau ) {X_6 \over 2}  \right], \nn \\
x \Del G (x) & = & -M^2 \pi \dsdtau \ \left\{
(X_8 + x X_{10}) + (X_9 + x X_{11}) 
+ \left( {\sig \over 2} - x \right) 
[(X_{12} + x X_{14}) + x (X_{15} + x X_{13})] \right. \nn \\
& & \left. - (x \sig - x^2 -\tau) \left[ X_{19} + X_{23} 
+ \left( {\sig \over 2} - x \right) X_{25} \right] \right\}, \nn \\
{\cal R}e \left( x \Del G_{3T} (x) \right) & = & 
M^2 \pi \dsdtau \ \left\{ .
(X_9 + x X_{11}) + \left( {\sig \over 2} - x \right) (X_{17} + x X_{19})
\right. \nn \\
& & \left. + \half (x \sig - x^2 -\tau) \left[ - (X_{15} + x X_{13})
- X_{19} + \left( {\sig \over 2} - x \right) X_{25} \right] \right\} , \nn \\
{\cal I}m \left( x \Del G_{3T} (x) \right) & = & 
M^2 \pi \dsdtau \ \left\{
- \left( {\sig \over 2} - x \right) (X_{16} + x X_{18})
- \half (x \sig - x^2 -\tau) \left[ X_{18} + 
\left( {\sig \over 2} - x \right) X_{24} \right] \right\}, \nn \\
{\cal R}e \left( x \Del H_{3T} (x) \right) & = & 
M^2 \pi \dsdtau \ \left\{(X_8 + x X_{10}) + 
\left( {\sig \over 2} - x \right) (X_{21} + x X_{23}) \right. \nn \\
& & \left. + \half (x \sig - x^2 -\tau)  \left[ -(X_{14}+xX_{13}) - X_{23} 
+ \left( {\sig \over 2} - x \right) X_{25} \right] \right\} , \nn \\
{\cal I}m \left( x \Del H_{3T} (x) \right) & = & 
M^2 \pi \dsdtau \ \left\{
- \left( {\sig \over 2} - x \right) (X_{20} + x X_{22}) 
- \half (x \sig - x^2 -\tau) 
\left[ X_{22} + \left( {\sig \over 2} - x \right) X_{24} \right] \right\}, \nn
\eea

\nin with the convention

\be
\left\{ d \sig d \tau \right\} 
\equiv d\sig d\tau \ \theta (x \sig - x^2 - \tau).
\ee



\end{document}